\documentclass[preprint,aps]{revtex4}
\usepackage{amssymb,epsf}

\begin{document}

\title{Slowly Rotating Black Holes in Einstein-Generalized Maxwell Gravity}
\author{S. H. Hendi\footnote{email address: hendi@mail.yu.ac.ir}}
\affiliation{Physics Department, College of Sciences, Yasouj
University, Yasouj
75914, Iran\\
Research Institute for Astrophysics and Astronomy of Maragha
(RIAAM), P.O. Box 55134-441, Maragha, Iran}

\begin{abstract}
In this paper, with considering the nonlinear electromagnetic
field coupled to Einstein gravity, we obtain the higher
dimensional slowly rotating charged black hole solutions. By use
of the fact that the temperature of the extreme black hole is
zero, we find that these solutions may be interpreted as black
hole solutions with inner (Cauchy) and outer (event) horizons
provide that the mass parameter $m$ is greater than an extremal
value $m_{ext}$, an extreme black hole if $m=m_{ext}$ and a naked
singularity otherwise. Also, we find that the asymptotic behavior
of the spacetime is not anti deSitter for the special values of
the nonlinearity parameter. Then, we compute the ADM mass,
electrical charge, temperature, entropy, angular momentum and
gyromagnetic ratio of the solutions. Calculations of the
electromagnetic field, electrical charge, entropy and temperature
showed that they are sensitive with respect to the changing of
nonlinearity parameter.
\end{abstract}

\pacs{04.40.Nr, 04.20.Jb, 04.70.Bw, 04.70.Dy}
\maketitle

Over the last decades a lot of attention has been focused on
rotating black hole solutions in presence of linear and nonlinear
electromagnetic field in the background. On one hand black holes
produced at colliders may in general have an electric charge as
well as other type of charges and therefore the study of charged
black hole becomes of great importance. On the other hand, the
question as to why the Planck and electroweak scales differ by so
many orders of magnitude remains mysterious. In recent years,
attempts have been made to address this hierarchy issue within the
context of theories with extra spatial dimensions. These subjects
motivate one to study on higher dimensional charged black hole
solutions. The first higher dimensional extensions of the
Schwarzschild solution have been obtained by Tangherlini
\cite{Tangherlini} and generalized by Myers and Perry
\cite{MyPer}. Also, in \cite{Fadeev}, higher dimensional charged
black hole solutions have been presented. This solution is the
generalization of the familiar Reissner-Nordstr\"{o}m solution for
a static and electrically charged black hole in ordinary general
relativity. Recently, many authors have introduced various classes
of charged rotating black hole solutions of Lovelock gravity and
investigate their thermodynamics \cite{ourchargepap}. Also,
properties of charged rotating black holes in Brans-Dicke theory
are investigated by some authors \cite{Pak}.

However the Kerr-Newman solution in higher dimensions, that is the
charged generalization of the Myers-Perry solution in the
Einstein-Maxwell gravity, it still remains to be found, but there
are many interesting paper focused on the slowly rotating black
holes \cite{Slowlyrot,Aliev1,Aliev2}.

From the other point of view, in the conventional, some
generalization of the matter sources to higher dimensions, one
essential property of the these sources is lost, namely, conformal
invariance. Electromagnetic field theory can be studied in a
special gauge which is conformally invariant, and firstly, has
been proposed by Eastwood and Singer \cite{EasSin}. In addition,
quantized Maxwell theory in a conformally invariant gauge and in
flat Euclidean 4-space have been investigated by Esposito
\cite{Esp} and one of the valuable work on nonlinear
electrodynamics have been done in \cite{Bronnikov}. Also, there
exists a generalized extension of the Maxwell action in higher
dimensions, if one uses the lagrangian of the U(1) gauge field in
the form \cite{HasMar1,HasMar2,HasMar3,Hendi1,HenRas,HendiPLB}
\begin{equation}
I_{CIM}=\alpha \int d^{d}x\sqrt{-g}\left( F_{\mu \nu }F^{\mu \nu
}\right) ^{s}  \label{non}
\end{equation}
where $F_{\mu \nu }=\partial _{\mu }A_{\nu }-\partial _{\nu
}A_{\mu }$ is the electromagnetic tensor field, $A_{\mu }$ is the
vector potential and $\alpha $ is a constant. It is
straightforward to show that the action (\ref{non}) is invariant
under conformal transformation ($g_{\mu \nu }\longrightarrow
\Omega ^{2}g_{\mu \nu }$ and $A_{\mu }\longrightarrow A_{\mu }$)
for $s=d/4$ and for $d=4$, the action (\ref{non}) reduces to the
Maxwell action as it should be \cite{HenRas}.

In the backdrop of the scenarios described so far it is therefore
worthwhile to study the slowly rotating black hole solutions in a
spacetime with negative cosmological constant in presence of
generalized electromagnetic field (\ref{non}).

The rest of the paper is organized as follows. We give a brief
review of the field equations of Einstein gravity in the presence
of generalized electromagnetic field in Sec. \ref{Field}. In Sec.
\ref{Solutions}, We present slowly rotating nonlinear charged
black hole solutions in Einstein gravity and then, we obtain mass,
charge, temperature, entropy, angular momentum, and gyromagnetic
ratio of the $(n+1)$-dimensional black hole solutions. Then, in
Sec. \ref{Conformally} we discuss about the special case of
nonlinear electromagnetic field, so-called conformally invariant
Maxwell field and investigate its properties. We finish our paper
with some concluding remarks.

\section{Field equations and solutions\label{Field}}

We consider the ($n+1$)-dimensional $(n\geq 3)$ spacetime in which
gravity is coupled to the nonlinear Maxwell field with an action
\begin{equation}
S=-\frac{1}{16\pi }\int_{\mathcal{M}}d^{n+1}x\sqrt{-g}\left[
R-2\Lambda +\left( \alpha F\right) ^{s}\right] -\frac{1}{8\pi
}\int_{\partial \mathcal{M}}d^{n}x\sqrt{-h}\Theta (h), \label{Act}
\end{equation}
where ${R}$ is the Ricci scalar, $\Lambda $ is the negative
cosmological constant, $F$ is the Maxwell invariant which is equal
to$\ F_{\mu \nu }F^{\mu \nu }$, $\alpha $ is a constant which we
should set it and $s$\ is a nonlinearity parameter. The last term
in Eq. (\ref{Act}) is the Gibbons-Hawking surface term and is
required for the variational principle to be well-defined. The
factor $\Theta $ represents the trace of the extrinsic curvature
for the boundary ${\partial \mathcal{M}}$ and $h$ is the induced
metric on the boundary. Varying the action (\ref{Act}) with
respect to the gravitational field $g_{\mu \nu }$ and the gauge
field $A_{\mu }$, yields
\begin{equation}
G_{\mu \nu }+\Lambda g_{\mu \nu }=T_{\mu \nu },  \label{FE1}
\end{equation}%
\begin{equation}
\partial _{\mu }\left[ \sqrt{-g}\left( \alpha F\right) ^{s-1}F^{\mu \nu }
\right] =0.  \label{FE2}
\end{equation}
In the presence of nonlinear electrodynamics field, the
energy-momentum tensor of Eq. (\ref{FE1}) is
\begin{equation}
T_{\mu \nu }=-2\left[ \alpha sF_{\mu \rho }F_{\nu }^{\rho }\left(
\alpha F\right) ^{s-1}-\frac{1}{4}g_{\mu \nu }\left( \alpha
F\right) ^{s}\right] , \label{TT}
\end{equation}
and for $s=1$ and $\alpha =-1$, Eqs. (\ref{FE1})-(\ref{TT}) reduce
to the standard Maxwell field coupled to Einstein gravity.

\section{The $(n+1)$-dimensional Slowly Rotating Nonlinear Charged Black
Holes\label{Solutions}}

For small rotation, we can solve Eqs. (\ref{FE1})-(\ref{TT}) to
first order in the angular momentum parameter $a$. Inspection of
the $(n+1)$-dimensional Kerr solutions shows that the only term in
the metric that changes to the first order of the angular momentum
parameter $a$ is $g_{t\phi }$. Therefore, for infinitesimal
angular momentum we assume the metric being of the following form
\begin{eqnarray}
ds^{2} &=&-U(r)dt^{2}+\frac{dr^{2}}{U(r)}-2aF(r)sin^{2}\theta
dtd\phi
\nonumber \\
&+&r^{2}(d\theta ^{2}+sin^{2}\theta d\phi ^{2}+cos^{2}\theta
d\Omega _{n-3}^{2})  \label{metric}
\end{eqnarray}
where $U(r)$ and $F(r)$ are functions of $r$, and $a$ is a
parameter associated with its angular momentum and $d\Omega
_{n-3}^{2}$ denotes the metric of an unit $(n-3)$-sphere. From
static cases we can consider the $t$ component of the Maxwell
equations in the form
\begin{equation}
F_{tr}=\xi h^{\prime }(r),  \label{Ftr}
\end{equation}
where prime denotes first derivative with respect to $r$ and $\xi
$ is a constant and may be fixed. In general, when we have
rotational parameter there is also a vector potential in the form
\begin{equation}
A_{\phi }=ah(r)sin^{2}\theta ,  \label{Ap}
\end{equation}
and one can show that for infinitesimal angular momentum, we have
\begin{equation}
F=\left( F_{\mu \nu }F^{\mu \nu }\right) =-2\left(
\frac{q}{(n-2)}h^{\prime }(r)\right) ^{2},
\end{equation}
and so the power Maxwell invariant, $\left( \alpha F\right) ^{s}$,
may be imaginary for positive $\alpha $, when $s$ is fractional.
Therefore we set $\alpha =-1$, to have real solutions without loss
of generality. By substituting Eq. (\ref{Ap}), the Maxwell fields
(\ref{Ftr}) and the metric (\ref{metric}) into the field equations
(\ref{FE1}) and (\ref{FE2}), we can obtain
\begin{eqnarray}
U(r) &=&1-\frac{2\Lambda r^{2}}{n(n-1)}-\frac{m}{r^{n-2}}+\Upsilon
(r),
\label{U(r)} \\
\Upsilon (r) &=&\left\{
\begin{array}{cc}
0, & s=0,1/2 \\
\frac{2^{n/2}q^{n}\ln r}{(n-2)^{n}r^{n-2}}, & s=n/2 \\
\frac{r^{2}(2s-1)^{2}}{(n-1)(2s-n)r^{2s(n-1)/(2s-1)}}\left( \frac{
2(2s-n)^{2}q^{2}}{(2s-1)^{2}(n-2)^{2}}\right) ^{s}, &
\text{otherwise}
\end{array}
\right.   \nonumber
\end{eqnarray}
\begin{eqnarray}
F(r) &=&\frac{n-2}{r^{n-2}}\left[ m-\Gamma (r)\right] ,  \label{F(r)} \\
\Gamma (r) &=&\left\{
\begin{array}{cc}
2\frac{(n-1)(s-1)}{(2s-n)}r^{n-2}, & s=0,1/2 \\
\left[ \frac{n}{2}+(n-2)\ln r\right] r^{n-2}+\frac{2^{n/2}q^{n}\ln
r}{(n-2)^{n}}, & s=n/2 \\
2\frac{(n-1)(s-1)}{(2s-n)}r^{n-2}+\frac{(2s-1)^{2}}{(n-1)(2s-n)}\left(
\frac{2(2s-n)^{2}q^{2}}{(n-2)^{2}(2s-1)^{2}}\right)
^{s}r^{(2s-n)/(2s-1)}, & \text{ otherwise}
\end{array}
\right.   \nonumber
\end{eqnarray}
\begin{equation}
h(r)=\left\{
\begin{array}{cc}
\text{cons}\tan \text{t,} & s=0,1/2 \\
q\ln r, & s=n/2 \\
qr^{(2s-n)/(2s-1)}, & \text{otherwise}
\end{array}
\right. .  \label{h(r)}
\end{equation}
In the above expressions, $q$ and $m$ appear as integration
constants and are related to the electrical charge and ADM
(Arnowitt-Deser-Misner) mass of the black hole, respectively. In
addition, these solutions reduce to slowly rotating
$4$-dimensional Kerr-Newman as $s=1$, $n=3$, $\Lambda =0$. Also,
in the linear case ($s=1$) and also as $\Lambda =0$, the solutions
reduce to the higher dimensional asymptotically flat slowly
rotating charged black hole solutions (see e.g. \cite{Aliev2}) and
we should set $\xi =-1/(n-2)$ for consistency. Likewise, the
presented solutions reduce to the solutions of obtained in
\cite{HasMar1,HasMar2} for $\Lambda =a=0$ (and $s=(n+1)/4$ for
\cite{HasMar1}). Also, this case ($s=(n+1)/4$) is correspond with
a class of $F(R)$ gravity in the absence of any matter source
\cite{HendiPLB}. In four dimension and in the absence of $\Lambda
$, the static solution ($a=0 $) is the special case of spherically
symmetric general solutions that presented in \cite{Bronnikov}.

In addition, for $s=n/2$ the charge term in metric function is
logarithmic, and the electromagnetic field is proportional to
$r^{-1}$, and in the other word this special solution is near to
BTZ solution in higher dimensions \cite{HendiBTZ} and
approximately, reduces to original static BTZ solution for
$3$-dimension ($n=2$) \cite{BTZ}.

Here, we want to investigate the effects of the nonlinearity on
the asymptotic behavior of the solutions.

a) $\Lambda <0$ :

It is worthwhile to mention that for $0<s<\frac{1}{2}$, the
asymptotic dominant term of Eq. (\ref{U(r)}) is fourth term and
the solutions of the slowly rotating nonlinear charged black hole
are not asymptotically AdS for negative $\Lambda $, but for the
cases $s<0$\ or $s>\frac{1}{2}$ (include of $s=\frac{n}{2}$), the
asymptotic behavior of solutions are the same as linear AdS case.
In spite of the fact that the physical implications of this effect
(different asymptotic behavior) is not exploited yet, but more
investigations about it can, independently, consider for future
works. Equations (\ref{Ftr})-(\ref{h(r)}) show that the
electromagnetic field is zero for the cases $s=0,\frac{1}{2}$, and
the metric functions do not possess a charge term and they
correspond to uncharged asymptotically AdS.

b) $\Lambda =0$:

For vanishing cosmological constant, one can show that the
presented solutions are asymptotically flat for all values of
nonlinearity parameter unless $\frac{-1}{n-3}\leqslant
s<\frac{1}{2}$.

At this point, it is worthwhile to investigate the causal
structure and physical properties of these solutions. One can show
that the Kretschmann scalar $R_{\mu \nu \lambda \kappa }R^{\mu \nu
\lambda \kappa }$ diverge at $r=0$ and it is finite for $r\neq 0$
and goes to zero as $r\rightarrow \infty $. So we find that there
is an essential singularity at $r=0$. Now, we look for the
existence of horizons. The horizons, if any exist, are given by
the zeros of the function $U(r)=(g_{rr})^{-1}$. Let us denote the
largest positive root of $U(r)=0$ by $r_{+}$.

Moreover we can obtain some information about causal structure by
considering the temperature of the black hole. By using the
definition of hawking temperature on the outer horizon $r_{+}$
which may be obtained through the definition of surface gravity
\begin{equation}
T_{+}=\frac{1}{2\pi }\sqrt{-\frac{1}{2}(\nabla _{\mu }\chi _{\nu
})(\nabla ^{\mu }\chi ^{\nu })}  \label{Tem1}
\end{equation}
where $\chi $ is the Killing vector $\partial _{t}$, we can write
\begin{eqnarray}
T_{+} &=&\frac{U^{\prime }(r)}{4\pi }=\frac{1}{4\pi }\left(
\frac{(n-2)m}{r_{+}^{n-1}}-\frac{4\Lambda r_{+}}{n(n-1)}+\Psi
\right) , \label{Temprature}
\\
\Psi  &=&\left\{
\begin{array}{cc}
0, & s=0,1/2 \\
\frac{2^{n/2}q^{n}}{(n-2)^{n}r_{+}^{n-1}}\left[ 1-(n-2)\ln
r_{+}\right] , &
s=n/2 \\
\frac{2(2s-1)(ns-3s+1)}{(n-1)(n-2s)}\left(
\frac{2(n-2s)^{2}q^{2}}{(n-2)^{2}(2s-1)^{2}}\right)
^{s}r_{+}^{(4s-2ns-1)/(2s-1)}, & \text{otherwise}
\end{array}
\right. .  \nonumber
\end{eqnarray}
Using the fact that the temperature of the extreme black hole is
zero, it is easy to show that the condition for having an extreme
black hole is that the mass parameter is equal to $m_{ext}$, which
is given as
\begin{equation}
m_{ext}=\frac{4\Lambda r_{+}^{n}}{n\left( n-1\right) \left(
n-2\right) } -\left\{
\begin{array}{cc}
0, & s=0,1/2 \\
\frac{2^{n/2}q^{n}}{(n-2)^{n+1}}\left[ 1-(n-2)\ln r_{+}\right] , & s=n/2 \\
\frac{2\,\left( 2\,s-1\right) \left( ns-3\,s+1\right) \left(
{\frac{2\left( n-2\,s\right) ^{2}{q}^{2}}{\left( n-2\right)
^{2}\left( 2\,s-1\right) ^{2}}} \right)
^{s}r_{+}^{(2\,s-n)/(2s-1)}}{\left( n-1\right) \left( n-2\right)
\left( n-2\,s\right) }, & \text{otherwise}
\end{array}
\right. .  \label{mext}
\end{equation}
One can show that the metric of Eqs. (\ref{metric})-(\ref{h(r)})
presents a slowly rotating black hole solution with two inner and
outer horizons provided that the mass parameter $m$ is greater
than $m_{ext}$, an extreme black hole for $m=m_{ext}$, and a naked
singularity otherwise.

In what follows we investigate the other conserved and
thermodynamics quantities. The entropy of the black hole typically
satisfies the so-called area law which states that the entropy of
the black hole is a quarter of the event horizon area
\cite{Becken}. This near universal law applies to almost all kinds
of black holes in Einstein gravity \cite{Hunter}. Since the area
of the event horizon does not change up to the linear order of the
rotating parameter $a$, we can easily show that the entropy of
black hole on the outer event horizon $r_{+}$ can be written as
\begin{equation}
S=\frac{V_{n-1}}{4}r_{+}^{n-1}  \label{Entropy}
\end{equation}
where $V_{n-1}$ represents the volume of constant curvature
hypersurface of an unit ($n-1$)-sphere, described by $d\Omega
_{n-1}^{2}=d\theta
_{1}^{2}+\sum\limits_{i=2}^{n-1}\prod\limits_{j=1}^{i-1}\sin
^{2}\theta _{j}d\theta _{i}^{2}$.

Next, we calculate the mass, angular momentum, electrical charge
and the gyromagnetic ratio of these rotating nonlinear charged
black holes which appear in the limit of slow rotation parameter.
The mass and angular momentum of the black hole can be calculated
through the use of the quasi-local formalism of the Brown and York
\cite{BY}. According to the quasilocal formalism, the quantities
can be constructed from the information that exists on the
boundary of a gravitating system alone. Such quasilocal quantities
will represent information about the spacetime contained within
the system boundary, just like the Gauss's law. In our case the
stress-energy tensor can be written as
\begin{equation}
T^{ab}=\frac{1}{8\pi }\left[ \Theta ^{ab}-\Theta \gamma
^{ab}\right] . \label{Stres}
\end{equation}
which is obtained by variation of the action (\ref{Act}) with
respect to $\gamma _{ab}$. To compute the angular momentum of the
spacetime, one should choose a spacelike surface $\mathcal{B}$ in
$\partial \mathcal{M}$ with metric $\sigma _{ij}$, and write the
boundary metric in ADM form
\begin{equation}
\gamma _{ab}dx^{a}dx^{a}=-N^{2}dt^{2}+\sigma _{ij}\left( d\varphi
^{i}+V^{i}dt\right) \left( d\varphi ^{j}+V^{j}dt\right) ,
\end{equation}
where the coordinates $\varphi ^{i}$ are the angular variables
parameterizing the hypersurface of constant $r$ around the origin,
and $N$ and $V^{i}$ are the lapse and shift functions,
respectively. When there is a Killing vector field $\mathcal{\xi
}$ on the boundary, then the quasilocal conserved quantities
associated with the stress-energy tensors of Eq. (\ref{Stres}) can
be written as
\begin{equation}
Q(\mathcal{\xi )}=\int_{\mathcal{B}}d^{n-1}x\sqrt{\sigma
}T_{ab}n^{a} \mathcal{\xi }^{b},
\end{equation}
where $\sigma $ is the determinant of the metric $\sigma _{ij}$,
$\mathcal{\xi }$ and $n^{a}$ are the Killing vector field and the
unit normal vector on the boundary $\mathcal{B}$, respectively.
For boundaries with timelike ($\xi =\partial /\partial t$) and
rotational ($\varsigma =\partial /\partial \varphi $) Killing
vector fields, one obtains the quasilocal mass and angular
momentum
\begin{equation}
M=\int_{\mathcal{B}}d^{n-1}x\sqrt{\sigma }T_{ab}n^{a}\xi
^{b}=\frac{V_{n-1}(n-1)m}{16\pi },  \label{mass}
\end{equation}
\begin{equation}
J=\int_{\mathcal{B}}d^{n-1}x\sqrt{\sigma }T_{ab}n^{a}\varsigma
^{b}=\frac{V_{n-1}ma}{8\pi }.  \label{Angmom}
\end{equation}
provided the surface $\mathcal{B}$ contains the orbits of
$\varsigma $. For $a=0$, the angular momentum vanishes, and
therefore $a$ is the rotational parameter of the nonlinear charged
black hole. The quasilocal mass that presented here, is the same
as ADM mass of calculated by Abbott and Deser \cite{AbbottDeser}.
Combining Eq. (\ref{mass}) with Eq. (\ref{Angmom}) we get
\begin{equation}
J=\frac{2Ma}{(n-1)}.  \label{J}
\end{equation}

Here, we can compute the electrical charge of the solutions. To
determine the electric field we should consider the projections of
the electromagnetic field tensor on special hypersurfaces. The
normal to such hypersurfaces for spacetimes with longitudinal
magnetic field is
\begin{equation}
u^{0}=\frac{1}{N},\text{ \ }u^{r}=0,\text{ \
}u^{i}=-\frac{N^{i}}{N},
\end{equation}
and the electric field is $E^{\mu }=g^{\mu \rho }F_{\rho \nu
}u^{\nu }$. The electric charge, ${Q}$ can be found by calculating
the flux of the electric field at infinity, yielding
\begin{equation}
Q=\frac{V_{n-1}}{4\pi }\times \left\{
\begin{array}{cc}
0, & s=0,\frac{1}{2} \\
\frac{2^{(n-4)/2}n}{(n-2)^{n-1}}q^{n-1}, & s=\frac{n}{2} \\
\frac{s}{2^{s}}\left[ \frac{2(2s-n)q}{(2s-1)(n-2)}\right] ^{2s-1},
& \text{otherwise}
\end{array}
\right. .  \label{charge}
\end{equation}

At last, we calculate the gyromagnetic ratio of this rotating
nonlinear charged black holes. One of the important subjects about
the $4$-dimensional charged black hole in the Einstein gravity is
that it can be assigned a gyromagnetic ratio g = 2 just like the
electron in Dirac theory. Here we want to know how does the value
of the gyromagnetic ratio change for slowly rotating nonlinear
charged black holes in higher dimensions. The magnetic dipole
moment for this slowly rotating black hole is
\begin{equation}
\mu =Qa.
\end{equation}
Therefore, the gyromagnetic ratio is given by
\begin{equation}
g=\frac{2QM}{\mu J}=n-1,  \label{Gyromagnet}
\end{equation}
which is the gyromagnetic ratio of the ($n+1$)-dimensional
Kerr-Newman black holes \cite{Aliev2}. Since both of angular
momenta and the magnetic dipole momenta of these black holes first
appear at the linear order in rotation parameter $a$, we have led
to the conclusion that the value of the gyromagnetic ratio remains
$g=n-1$. Also, we find that, the nonlinearity of electromagnetic
field does not change the gyromagnetic ratio of the rotating black
hole.

\section{Conformally invariant electromagnetic field\label{Conformally}}

As one can see the clue of the conformal invariance of Maxwell
source lies in the fact that we have considered power of the
Maxwell invariant, $F=F_{\mu \nu }F^{\mu \nu }$
\cite{HasMar1,HasMar2,HasMar3,HenRas}. Here we want to justify the
nonlinearity parameter $s$, such that the electromagnetic field
equation be invariant under conformal transformation ($g_{\mu \nu
}\longrightarrow \Omega ^{2}g_{\mu \nu }$ and $A_{\mu
}\longrightarrow A_{\mu }$). The idea is to take advantage of the
conformal symmetry to construct the analogues of the four
dimensional Reissner-Nordstr\"{o}m solutions in higher dimensions.
It is easy to show that for Lagrangian in the form $L(F)$ in
$(n+1)$-dimensions, $T_{\mu }^{\mu }\propto \left[
F\frac{dL}{dF}-\frac{n+1}{4}L\right] $; so $T_{\mu }^{\mu }=0 $
implies $L(F)=Constant\times F^{(n+1)/4}$. For our case
$L(F)\propto F^{s}$, and subsequently, $s=(n+1)/4$. It is
worthwhile to mention that Since $n\geq 3$ and therefore $s\geq
1$, one can show that the slowly rotating black holes with
conformally invariant Maxwell source are asymptotically AdS in
arbitrary dimensions. In this case the functions $U(r)$, $F(r)$
and $h(r)$ reduce to
\begin{equation}
U(r)=1-\frac{2\Lambda
r^{2}}{n(n-1)}-\frac{m}{r^{n-2}}-\frac{2^{(n-3)/4}}{
r^{n-1}}\left( \frac{q}{n-2}\right) ^{(n+1)/2},
\end{equation}
\begin{equation}
F(r)=\frac{(n-2)}{r^{n-2}}\left[
m+(n-3)r^{n-2}+\frac{2^{(n-3)/4}}{r}\left( \frac{q}{n-2}\right)
^{(n+1)/2}\right]
\end{equation}
\begin{equation}
h(r)=\frac{1}{r}
\end{equation}
and therefore $F_{tr}\propto r^{-2}$ in arbitrary dimensions.

\section{Summary and Conclusion\label{Conclusion}}

In this paper, we presented higher dimensional slowly rotating
nonlinear charged black hole solutions in Einstein gravity in the
presence of negative cosmological constant. We discarded any terms
involving $a^{2}$ or higher power in $a$. The nonlinearity does
not change to $O(a)$ and $A_{\phi }$ is the only component of the
vector potential that change to $O(a)$. These solutions may be
interpreted as black hole solutions with inner (Cauchy) and outer
(event) horizons provided that the mass parameter $m$ is greater
than an extremal value given by Eq. (\ref{mext}), an extreme black
hole if $m=m_{ext}$ and a naked singularity otherwise. We showed
that the vector potential and the metric functions are logarithmic
form for special case, $s=n/2$. This case is analogous with the
BTZ solutions. Also, we presented the effects of nonlinearity on
the solutions and discussed about the asymptotic behavior of them.
For the special choices of the nonlinearity parameter $s$, these
solutions do not have asymptotic AdS behavior. Calculations of the
electromagnetic field, electrical charge and temperature showed
that they are sensitive with respect to the changing of
nonlinearity parameter. The expressions of the mass, temperature,
and entropy of the black hole solution show that they do not
change up to the linear order of the angular momentum parameter
$a$. One may think there is not any rotational effect at infinity
for the linear order in rotation parameter $a$.

Then, we obtained the ADM mass $M$, the angular momentum $J$ and
the gyromagnetic ratio $g$ of the black hole and found that they
do not depend on the nonlinearity parameter $s$. Finally, we
adjusted the nonlinearity parameter, $s$, such that, in higher
dimensions, the electromagnetic field equations be invariant under
conformal transformations and then $F_{tr}\propto r^{-2}$.

\begin{acknowledgements}
This work has been supported financially by Research Institute for
Astronomy and Astrophysics of Maragha.
\end{acknowledgements}

\end{document}